\documentclass[twocolumn,showpacs,preprintnumbers,amsmath,amssymb]{revtex4}

\usepackage{graphicx}
\usepackage{dcolumn}
\usepackage{bm}
\draft

\begin{document}
\def\bvv{{\hbox{\boldmath $v$}}}

\title{Pion interferometry for hydrodynamical expanding source with a 
finite baryon density}

\author{W. N. Zhang$^{1,2}$}
\author{M. J. Efaaf$^1$}
\author{Cheuk-Yin Wong$^{2,3}$}
\author{M. Khaliliasr$^1$}

\affiliation{
$^1$Department of Physics, Harbin Institute of Technology, 
Harbin, 150006, P. R. China\\
$^2$Physics Division, Oak Ridge National Laboratory, Oak Ridge, TN
37831, U.S.A.\\
$^3$Department of Physics, University of Tennessee, Knoxville, TN
37996, U.S.A.
}

\date{\today}

\begin{abstract}

We calculate the two-pion correlation function for an expanding hadron
source with a finite baryon density. The space-time evolution of the
source is described by relativistic hydrodynamics and the HBT radius
is extracted after effects of collective expansion and multiple
scattering on the HBT interferometry have been taken into account,
using quantum probability amplitudes in a path-integral formalism.  We
find that this radius is substantially smaller than the HBT radius
extracted from the freeze-out configuration.

\end{abstract}


\maketitle

The Bose-Einstein correlation of identical bosons produced in
high-energy heavy-ion collisions, also known as the HBT effect, is an
important tool for the study of the space-time structure of the
emitting source \cite{Won94}.  As the source expands, cools, and 
freezes out, it is important to know what source distribution the HBT 
interferometry measures. Is it the freeze-out source distribution, the 
initial source distribution, or the initial source distribution
modified by absorption and expansion?  The conventional viewpoint is
that the HBT interferometry measures the freeze-out configuration
because rescattering of source particles are assumed to lead to a
chaotic source.  However, the extracted experimental HBT radii are
insensitive to the collision energy, and $R_{\rm out}/R_{\rm side}
\approx 0.9- 1.1$ at RHIC \cite{PHE02,STA01}.  These general features cannot 
be understood within the conventional viewpoint \cite{Mol04,Pra03,Hei02}.

The validity of the conventional assumption on HBT is recently
questioned as it was pointed out that because HBT is purely a
quantum-mechanical phenomenon, the effects of rescattering and
collective dynamics must be investigated within a quantum-mechanical
context \cite{Won03}.  As fully quantum treatment of heavy-ion collisions 
is beyond the scope of our present investigation, we shall describe the 
gross dynamics of the hadron system with a finite baryon density by 
classical hydrodynamics.  In this letter, we follow Wong's work 
\cite{Won03} and take into account the effects of collective expansion and 
multiple scattering by constructing the quantum probability amplitude using 
Glauber's multiple scattering model and the trajectories of test particles 
in the path-integral framework.

The hydrodynamics and the composition of the fluid depend on the
collision energy.  We study in this work the dynamics of heavy-ion
collisions in the energy range corresponding to about 10 GeV per
nucleon beam energy in fixed target collisions in which nucleons and 
deltas play an important role and pions
are produced dominantly from deltas and absorbed by nucleons.

Relativistic hydrodynamics has been extensively applied to high-energy
heavy-ion collisions \cite{Kol03}.  The energy momentum tensor of a 
thermalized fluid cell in the center-of-mass frame is \cite{Kol03}
\begin{equation}
\label{tensor}
T^{\mu \nu} (x) = \big [ \epsilon(x) + p(x) \big ] u^{\mu}(x) u^{\nu}(x) 
- p(x) g^{\mu \nu} \, ,
\end{equation}
where $x$ is the space-time coordinate, $\epsilon$, $p$, and
$u^{\mu}=\gamma (1,\bvv)$ are respectively the energy density,
pressure, and 4-velocity of the cell, and $g^{\mu \nu}$ is the metric
tensor.  The local conservation of energy and momentum can be expressed by
\begin{equation}
\label{paremt}
\partial_{\mu} T^{\mu \nu}(x)=0 , ~~~~(\nu =0,1,2,3) .
\end{equation}
The conservation of baryon number gives
\begin{equation}
\label{current}
\partial_{\mu} j^{\mu}(x)=0 ,
\end{equation}
where $j^{\mu}=n_b(x) u^{\mu}$ is the four-current-density of baryon 
($n_b$ is baryon density). 
In the thermalized cell, the density number $n_i$ of the 
particle species $i$, the energy density $\epsilon_i$, the pressure
$p_i$, and the local velocity of sound $c_s$ can be obtained as a function
of local temperature $T(x)$ and local chemical potential $\mu_i (x)$
by assuming an ideal hadron gas.  For simplicity, we consider a 
hadronic gas consisting of nucleons, $\Delta (1232)$, and pions with 
spherical geometry as in Ref. \cite{Ris9896}.  For spherical geometry
Eqs. (\ref{paremt}) and (\ref{current}) become \cite{Ris9896}
\begin{eqnarray} 
\label{eqe}
\partial_t E + \partial_r [(E+p)v] & = &
- F \, , \\
\label{eqm}
\partial_t M + \partial_r (Mv+p) & = &
- G \, , \\
\label{eqr}
\partial_t N + \partial_r (Nv) & = &
- U \, , 
\end{eqnarray}
where $E \equiv T^{00}$, $M \equiv T^{or}$, $N \equiv n_b \gamma$, 
\begin{eqnarray} 
\label{FGR}
F = \frac{2 v}{r} (E+p),~~~~G = \frac{2 v}{r} M ,~~~~
U = \frac{2 v}{r} N .
\end{eqnarray}
In order to bring out the salient features of the effects under 
consideration, we shall carry out model calculations for some idealized  
space-time configurations.  In these model calculations, we assume that 
the initial velocity $v(0,r)=0$, and the initial energy density is a 
Gaussian distribution 
$\epsilon(0,r)=\epsilon_0 \,e^{-r^2/2R_0^2}$ \cite{Mia02}. 
In a local equilibrium cell, we have $\mu_{\pi}=0$, $\mu_N =\mu_{\Delta}
\equiv \mu_B$.  Using the HLLE scheme \cite{Ris9896} and with the relations 
of $\epsilon(T(x),\mu_B(x))$, $p(T(x),\mu_B(x))$, $n_b(T(x),\mu_B(x))$, 
and $c_s(T(x),\mu_B(x))$, we can get the solution of the hydrodynamical 
equations for $F=G=U=0$.  Then, using the Sod's operator splitting 
method \cite{Ris9896}, we can obtain the solution for Eqs. (\ref{eqe}), 
(\ref{eqm}), and (\ref{eqr}).

Figure 1 shows the temperature and velocity for a spherical expanding
source with $\epsilon_0=0.5$ GeV/fm$^3$ and $R_0 = 4.0$ fm. The
corresponding initial temperature and baryon density at $r=0$ are
$T_0=137$ MeV and $n_{b0}=0.36$ fm$^{-3}$, respectively.  They are 
obtained by using the state of equation of mixed ideal gas and $\epsilon_0$. 
The temperature $T_0=137$ MeV is lower than the estimated critical temperature, 
$T_c = 160$ MeV, of the transition between quark-gluon plasma and hadronic 
gas \cite{Ris9896}. 

\begin{figure}
\includegraphics{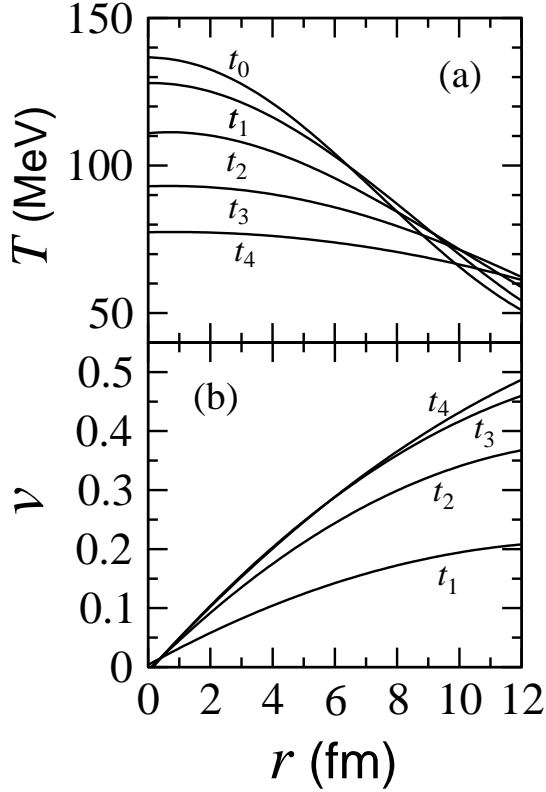}
\caption{\label{fig:tv} Temperature (a) and velocity (b) profiles for the 
expanding source at $t_n = 3n$ fm/c.}
\end{figure}

The two-particle Bose-Einstein correlation function is defined as 
the ratio of the two-particle momentum distribution $P(k_1,k_2)$ to 
the product of the single-particle momentum distribution $P(k_1)
P(k_2)$.  For an expanding source, $P(k)$ and $P(k_1,k_2)$ can be 
expressed as \cite{Won03}
\begin{eqnarray}
\label{pk1}
P(k) = \int d^4x \, e^{-2\,{\cal I}{m}\,{\bar \phi}_s(x)}
\rho(x) A^2(\kappa(x),x) \,,
\end{eqnarray}
\vspace*{-0.4cm}
\begin{eqnarray}
\label{pk12}
\!\!\!\!\!\!\!\!\!P(k_1,k_2)\!\!\!
&=& \!\!\!\! \int \!\! d^4x_1 d^4x_2\, e^{-2\,{\cal I}{m}\,{\bar \phi}_s(x_1)} 
e^{-2\,{\cal I}{m}\,{\bar \phi}_s(x_2)} 
\nonumber \\
&\times& \!\!\!\rho(x_1) \rho(x_2) 
|\Phi(\kappa_1 \kappa_2: x_1 x_2 \rightarrow x_{d1} 
x_{d2})|^2 ,
\end{eqnarray}
where $A(\kappa (x),x)$ is the magnitude of the amplitude for
producing a pion with momentum $\kappa$ at $x$, which is proportional to
the Bose-Einstein distribution characterized by the local temperature in the 
local frame of the cell at $x$.  $\rho(x)$ is the pion-source 
density, which includes a
primary source and the secondary source from $\Delta$ decay,
\begin{eqnarray}
\label{rho}
\rho(x) &=& \rho^{\rm prim}(x) + \rho^{\rm second}(x)\nonumber\\
&=&n_{\pi}(x) \, \delta(t) + \Gamma \, n_{\Delta}(x) \,,
\end{eqnarray}
where $\Gamma=120$ MeV is the width of $\Delta$. $e^{-2\,{\cal
I}{m}\,{\bar \phi}_s(x)}$ is the absorption factor due to multiple
scattering \cite{Won03,Won04}, 
\begin{equation}
\label{absor}
e^{-2\,{\cal I}{m}\,{\bar \phi}_s(x)}=\exp\bigg(- \int_{x}^{ x_f} 
\sigma_{\rm abs} (\sqrt{s_{\pi N}}) \, n_N(x) \, d l \bigg) ,
\end{equation}
where $\sigma_{\rm abs} (\sqrt{s_{\pi N}})$ is the absorption cross section 
of $\pi + N \rightarrow \Delta$ at the center-of-mass energy $\sqrt{s_{\pi 
N}}$ and $d l$ is the spatial line element along the path of particle
propagation. In Eq. (\ref{pk12}), $\Phi (\kappa_1 \kappa_2 : x_1 x_2
\rightarrow x_{d1} x_{d2})$ is the wave function for two-particles
produced at $x_1$ and $x_2$ with momenta $\kappa_1 (x_1)$ and
$\kappa_2 (x_2)$, and detected at $x_{d1}$ or $x_{d2}$ with momenta
$k_1$ and $k_2$, respectively,
\begin{eqnarray}
\label{PHI}
&&\!\!\!\!\!
\Phi (\kappa_1 \kappa_2 : x_1 x_2 \rightarrow x_{d1} x_{d2})
\nonumber\\
&&= { 1 \over \sqrt{2}} \biggl \{\! A(\kappa_1(x_1), x_1) 
A(\kappa_2(x_2) x_2)
\nonumber\\
&&\times \exp\Big[-i\int_{x_1}^{x_{f1}} \!\! \kappa_1(x') \cdot dx' 
-i  {k}_1 \cdot (x_{d1}-x_{f1})\Big] 
\nonumber\\
&&\times\exp\Big[-i\int_{x_2}^{x_{f2}} \!\!\kappa_2(x') \cdot dx'
-i  {k}_2 \cdot (x_{d2}-x_{f2})\Big] 
\nonumber\\
&&+ ~A(\kappa_1(x_2), x_2) A(\kappa_2(x_1), x_1)
\nonumber\\
&&\times \exp\Big[-i\int_{x_2}^{x'_{f2}} \!\! \kappa_1(x') 
\cdot dx' -i  {k}_1 \cdot (x_{d1}-x'_{f2}) \Big] 
\nonumber\\
&&\times \exp\Big[-i\int_{x_1}^{x'_{f1}} \!\! \kappa_2(x') 
\cdot dx' - i {k}_2 \cdot (x_{d2}-x'_{f1}) \Big] \bigg\} ,~~~~~~
\end{eqnarray}
where $x_{f1}$ and $x_{f2}$ are the freeze-out points corresponding to
the particles 1 and 2 produced at $x_1$ and $x_2$ and detected at
$x_{d1}$ and $x_{d2}$, respectively. $x'_{f1}$ and $x'_{f2}$ are the
freeze-out points corresponding to the particles 1 and 2 produced at
$x_1$ and $x_2$ and detected at $x_{d2}$ and $x_{d1}$, respectively.
As we consider mainly two particles whose momenta are nearly
parallel, we approximately have $x'_{f1}= x_{f1}$ and $x'_{f2}
=x_{f2}$ \cite{Won03,Won04a}.  From Eqs. (\ref{pk1}), (\ref{pk12}), and 
(\ref{PHI}) the correlation function $C(k_1,k_2)=P(k_1,k_2)/P(k_1)P(k_2)$ 
can be written as \cite{Won03} 
\begin{eqnarray}
\label{ck12m}
C(k_1,k_2) &=& 1 + \biggl \vert \int
d^4x \, e^{ i(k_1-k_2) \cdot x +i\phi_c(x, k_1 k_2 )}
\nonumber\\
& &\times e^{-2 ~{\cal I}{m}~ {\bar \phi}_s(x)}
\rho_{\rm eff} (x; k_1 k_2) \biggr \vert ^2 \, ,
\end{eqnarray}
where $\rho_{\rm eff}$ is the effective density
\begin{eqnarray}
\label{effrf}
\rho_{\rm eff} (x; k_1 k_2)=
{ \sqrt{f_{\rm init}(\kappa_1 (x),x)} \sqrt {f_{\rm init}(\kappa_2 (x),x)}
\over \sqrt{ P(k_1) P(k_2) } } \,,
\end{eqnarray}
\begin{eqnarray}
\label{finit}
f_{\rm init}(\kappa (x),x) =
\rho(x) A^2(\kappa (x),x)  \,,
\end{eqnarray}
is the phase space distribution of particle sources, and
\begin{eqnarray}
\label{phic}
\phi_c (x, k_1 k_2)=
 - \!\int_{x}^{x_{f}}\!
\{[\kappa_1(x')-\kappa_2(x')]-[k_1-k_2] \}\cdot dx'.
\!\!\!\! \nonumber\\
\end{eqnarray}
It can be seen that the two-pion correlation function is related to
the phase space distribution of the pion production source, modified
by an absorption factor arising from multiple scattering and a phase
factor $\phi_c$ from the collective expansion.

Knowing the hydrodynamical solution as the space-time variations of
density, velocity, and thermodynamical functions of the fluid cells,
we can construct the trajectories of a pair of test pions which start
initially at $x_1$ and $x_2$ and emerge finally with momenta $k_1$ and
$k_2$ at freeze-out.  The knowledge of the space-time trajectories of
the pair of pions allows one to calculate the path integrals in the
wave function $\Phi$ of Eq.\ (\ref{PHI}) and the function $P(k_1,k_2)$
and $P(k)$ after summing over all source elements in Eqs.\ (\ref{pk1})
and (\ref{pk12}).  The correlation function $C(q)$ of the relative
momentum of the two particles, $q=|{\bf k_1}-{\bf k_2}|$, can be
constructed from $P(k_1,k_2)$ and $P(k_1)P(k_2)$ by integrating over
the average momentum $({\bf k_1}+{\bf k_2})/2$ \cite{Zha93}. 
The HBT radius $R$ can then be extracted by parameterizing the correlation
function $C(q)$ as
\begin{equation}
\label{cq}
C(q)=1+\lambda e^{-q^2 R^2} \,.
\end{equation}

\begin{figure}
\includegraphics{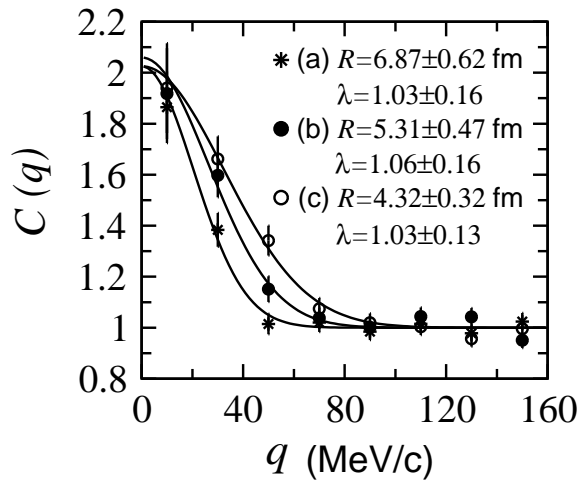}
\caption{\label{fig:hbt} Two-pion correlation functions $C(q)$ for the
cases (a) freeze-out, (b) present formulation including absorption of
pions by nucleons, and (c) without absorption.}
\end{figure}

In our model calculation, we shall consider the freeze-out configuration to 
be characterized by a freeze-out temperature of $T_f=0.5 T_0$.  We calculate 
first the correlation function $C(q)$ for the case (a) when the detected 
pions are assumed to originate only from the freeze-out configuration.  
This is the configuration considered in the conventional
description of intensity interferometry with collective expansion where 
one assumes a chaotic source at freeze-out due to random pion
rescattering.

We calculate next the case (b) using the present formalism to take
into account the effects of multiple scattering and collective
expansion with quantum probability amplitudes in a path-integral
framework.  We use the experimental absorption cross 
section \cite{Pdg02} represented by the model of the 
absorption of a pion by a nucleon forming a delta resonance.  For further 
comparison, we calculate $C(q)$ for the case (c) in which the detected 
pions originate from the source under the hydrodynamical expansion with no
absorption.

Figure 2 shows results of the two-pion interferometry for these
cases.  The conventional freeze-out configuration [case (a)]
leads to an extracted HBT radius of 6.87 fm.  When there is no
absorption of pions [case (c)], the HBT radius of 4.32 fm is small.
The HBT radius obtained by using the present formalism [case (b)]
leads to an HBT radius of 5.31 fm which lies between the radii of case
(a) and case (c). The extracted HBT radius depend on the description
of the pion source and absorption.

We conclude that for a pion source of the type considered, the HBT
radius extracted from the freeze-out configuration is substantially
greater than the radius obtained when effects of multiple scattering
and collective expansion on the HBT interferometry have been properly
taken into account using quantum probability amplitudes in a
path-integral formalism.  As the latter description is a more
realistic description of the HBT interferometry, the conventional
assumption that the HBT interferometry measures the distribution of
the freeze-out configuration is therefore subject to question.

We have carried out model calculations for some idealized configurations 
in order to illustrate the main features of the effects under consideration. 
These features will remain the same for more realistic space-time 
configurations.  In this paper the hadronic gas is taken to be an ideal gas.  
For a more realistic case, one should consider the volume correction 
\cite{Bar89,Zha00}, which will lead to a slightly lower temperature and 
lower density of hadronic gas but will hardly affect the main features of 
the HBT interferometry considered here.  The freeze-out temperature $T_f=
0.5T_0$ considered in this paper is lower than the freeze-out temperatures 
$T_f=0.7T_c$ and $0.9T_c$ ($T_c=160$ MeV) considered for a zero net baryon 
density case \cite{Ris9896} because the cross section between pion and baryon 
is greater than that between pions.  The HBT radii of the three cases 
considered in this paper will increase when the freeze-out temperature 
decreases but the main feature concerning the relative magnitudes of the 
HBT radii is unaltered. 

WNZ would like to thank Drs. T. Barnes, V. Cianciolo, and G. Young for
their kind hospitality at Oak Ridge National Laboratory.  This
research was supported by the National Natural Science Foundation of
China under Contract No.10275015 and by the Division of Nuclear
Physics, US DOE, under Contract No. DE-AC05-00OR22725 managed by
UT-Battle, LC.


\begin{thebibliography}{99}

\bibitem{Won94}
For a review see Chapter 17 of C. Y. Wong, Introduction to High-Energy
Heavy-Ion Collisions, World Scientific Publishing Company, 1994;
U. A. Wiedemann, U. Heinz, Phys. Rept. {\bf 319}, 145 (1999); R. M. Weiner,
Phys. Rept. {\bf 327}, 249 (2002).
                                                                                
\bibitem{PHE02}
K. Adcox et al., Phys. Rev. Lett. {\bf 88}, 192302 (2002).
                                                                                
\bibitem{STA01}
C. Adler et al., Phys. Rev. Lett. {\bf 87}, 082301 (2001).
                                                                                
\bibitem{Mol04}
D. Moln${\rm \acute{a}}$r and M. Gyulassy, Phys. Rev. Lett. {\bf 92},
052301 (2004).
                                                                                
\bibitem{Pra03}
S. Pratt, Nucl. Phys. A {\bf 715}, 389c (2003).
                                                                                
\bibitem{Hei02}
U. Heinz and P. Kolb, Nucl. Phys. A {\bf 702}, 269 (2002).
                                                                                
\bibitem{Won03}
Cheuk-Yin Wong, J. Phys. G {\bf 29}, 2151 (2003).
                                                                                
\bibitem{Kol03}
P. Kolb and U. Heinz, nucl-th/0305084.
                                                                                
\bibitem{Ris9896}
D. H. Rischke, M. Gyulassy, Nucl. Phys A {\bf 608}, 479 (1996);
D. H. Rischke, nucl-th/9809044.
                                                                                
\bibitem{Mia02}
H. Miao, Z. Ma, C. Gao, Commun. Theor. Phys. {\bf 38}, 698 (2002).
                                                                                
\bibitem{Won04}
Cheuk-Yin Wong, and R. Glauber, to be submitted.
                                                                                
\bibitem{Won04a}
Cheuk-Yin Wong, J. Phys. G {\bf 30}, S1053 (2004). 
                                                                                
\bibitem{Zha93}
W. N. Zhang, Y. M. Liu, S. Wang, Q. J. Liu, J. Jiang, D. Keane, Y. Shao, 
S. Y. Chu, and S. Y. Fung, Phys. Rev. C {\bf 47}, 795 (1993).
                                                                                
\bibitem{Pdg02}
K. Hagiwara et al., Phys. Rev. D {\bf 66}, 010001 (2002).

\bibitem{Bar89}
S. P. Baranov and L. V. Fil'kov, Z. Phys. C {\bf 44}, 227 (1989). 

\bibitem{Zha00}
W. N. Zhang, G. X. Tang, X. J. Chen, L. Huo, Y. M. Liu, and S. Zhang, 
Phys. Rev. C {\bf 62}, 044903 (2000). 

\end{thebibliography}
\end{document}